\documentclass[aps,prl,twocolumn,10pt,nofootinbib,nobibnotes,longbibliography]{revtex4-1}

\usepackage{amssymb}
\usepackage{graphicx}
\usepackage{amsmath}
\usepackage{hyperref}
\usepackage{subfigure}
\usepackage{multirow}
\usepackage{setspace}
\usepackage{verbatim}
\usepackage{float}
\usepackage{color}
\usepackage{ulem}
\usepackage[utf8]{inputenc}
\usepackage[table,xcdraw]{xcolor}

\begin{document}

\title{Constraints on holographic QCD phase transitions from PTA observations}

\author{Song He$^{1,2,7}$}
\email{hesong@jlu.edu.cn}

\author{Li Li$^{3,4,6}$}
\email{liliphy@itp.ac.cn}

\author{Sai Wang$^{5,6}$}
\email{wangsai@ihep.ac.cn}

\author{Shao-Jiang Wang$^{3}$}
\email{schwang@itp.ac.cn}

\affiliation{$^1$ Center for Theoretical Physics and College of Physics, Jilin University, Changchun 130012, China}

\affiliation{$^2$ School of Physical Science and Technology, Ningbo University, Ningbo, 315211, China}

\affiliation{$^3$ CAS Key Laboratory of Theoretical Physics, Institute of Theoretical Physics, Chinese Academy of Sciences, Beijing 100190, China}

\affiliation{$^4$ School of Fundamental Physics and Mathematical Sciences, Hangzhou Institute for Advanced Study, \\University of Chinese Academy of Sciences (UCAS), Hangzhou 310024, China}

\affiliation{$^5$ Theoretical Physics Division, Institute of High Energy Physics, Chinese Academy of Sciences, Beijing 100049, China}

\affiliation{$^6$ School of Physical Sciences, University of Chinese Academy of Sciences (UCAS), Beijing 100049, China}

\affiliation{$^7$ Max Planck Institute for Gravitational Physics (Albert Einstein Institute), Am M\"uhlenberg 1, 14476 Golm, Germany}

\begin{abstract}
The underlying physics of QCD phase transition in the early Universe remains largely unknown due to its strong-coupling nature during the quark-gluon plasma/hadron gas transition, yet a holographic model has been proposed to quantitatively fit the lattice QCD data while with its duration of the first-order phase transition (FoPT) left undetermined. At specific baryon chemical potential, the first-order QCD phase transition agrees with the observational constraint of baryon asymmetry. It, therefore, provides a scenario for phase transition gravitational waves (GWs) within the Standard Model of particle physics. If these background GWs could contribute dominantly to the recently claimed common-spectrum red noise from pulsar timing array (PTA) observations, the duration of this FoPT can be well constrained, and the associated primordial black holes are still allowed by current observations.
\end{abstract}
\maketitle

\textit{\textbf{Introduction.}---}
Recently, independent evidence for detecting a gravitational-wave (GW) background around the nano-Hz band has been reported by different observations using pulsar timing array (PTA)~\cite{Xu:2023wog, NANOGrav:2023gor, EPTA:2023sfo, Reardon:2023gzh}, among which the Chinese PTA Data Release I (CPTA DR1)~\cite{Xu:2023wog} has found the highest statistical significance ($4.6\sigma$) for the Hellings–Downs correlation curve~\cite{Hellings:1983fr}, while the North American Nanohertz Observatory for Gravitational Waves 15-year data (NANOGrav 15yr)~\cite{NANOGrav:2023gor} has put strong constraints on the excluded parameter spaces for various cosmological sources~\cite{NANOGrav:2023hvm} (see also~\cite{Madge:2023dxc}) when their GWs significantly exceed the NANOGrav
signal. Nevertheless, although the 10.3-year subset of European Pulsar Timing Array second data release (EPTA DR2)~\cite{EPTA:2023sfo} based on modern observing systems renders a 15 times larger Bayes factor of GW background detection compared to that of the full 24.7-year EPTA data set, its inferred spectrum is in mild tension with the common signal measured in the full data set. Similarly, the first half of the Parkes Pulsar Timing Array third data release (PPTA DR3)~\cite{Reardon:2023gzh} yields an upper limit on the inferred common-spectrum amplitude in tension with that from the complete data.

However, if the signal
is indeed genuine, we are in a position to search for other GW sources (for example, cosmic inflation~\cite{Vagnozzi:2023lwo,Borah:2023sbc,Datta:2023vbs,Niu:2023bsr,Choudhury:2023kam,Jiang:2023gfe,Ben-Dayan:2023lwd}, scalar-induced GWs~\cite{Wang:2023ost,Franciolini:2023pbf,Bian:2023dnv,Inomata:2023zup,HosseiniMansoori:2023mqh,Yi:2023mbm,Liu:2023ymk,Abe:2023yrw,Jin:2023wri,Wang:2023sij,Balaji:2023ehk,Zhu:2023gmx,Liu:2023pau}, phase transitions~\cite{Addazi:2023jvg,Athron:2023mer,Fujikura:2023lkn,Han:2023olf,Franciolini:2023wjm,Bian:2023dnv,Jiang:2023qbm,Ghosh:2023aum,Xiao:2023dbb,Li:2023bxy,DiBari:2023upq,Cruz:2023lnq,Wu:2023hsa,Du:2023qvj,Gouttenoire:2023bqy,Ahmadvand:2023lpp,Wang:2023bbc}, domain walls~\cite{Bai:2023cqj,Kitajima:2023cek,Bian:2023dnv,Blasi:2023sej,Gouttenoire:2023ftk,Barman:2023fad,Lu:2023mcz,Wu:2023hsa,Babichev:2023pbf,Zhang:2023nrs,Ge:2023rce}, cosmic strings~\cite{Ellis:2023tsl,Wang:2023len,Bian:2023dnv,Lazarides:2023ksx,Wu:2023hsa,Servant:2023mwt,Antusch:2023zjk,Yamada:2023thl}, and ultralight dark matter~\cite{Aghaie:2023lan}, to name just a few, as long as they preserve the causality~\cite{Giblin:2014gra}) in addition to the conventional background from inspiraling supermassive black hole binaries (SMBHBs), even though the SMBHB background itself might also call for better modeling from unknown environmental effects~\cite{NANOGrav:2023hfp,Ellis:2023dgf,Shen:2023pan,Bi:2023tib,Ghoshal:2023fhh,Depta:2023qst,Gouttenoire:2023nzr,Hu:2023oiu}.

The cosmological quantum chromodynamics (QCD) phase transition holds significant implications as a potential source for a stochastic GW background if it is of the first order. However, constraints imposed by primordial element abundances and the cosmic microwave background have led to a stringent limitation on the baryon asymmetry $\eta_B\equiv n_B/s$, where $n_B$ and $s$ denote the baryon number density and entropy density, respectively \cite{dodelson2003modern}. The observed value~\cite{Fields:2019pfx}, $\eta_B\approx 10^{-10}$, has led to the prevailing belief that a cosmological first-order QCD phase transition does not occur within the Standard Model of particle physics. Hence, numerous QCD model buildings beyond the Standard Model have been proposed to introduce a first-order phase transition (FoPT) (e.g., see specifically Refs.~\cite{Boeckel:2009ej,Boeckel:2011yj,Schwaller:2015tja,Aoki:2017aws,Iso:2017uuu,Bai:2018dxf,Lu:2022yuc,Sagunski:2023ynd,Salvio:2023ynn} and most recent review~\cite{Athron:2023xlk}).

\begin{figure*}
\includegraphics[width=0.8\textwidth]{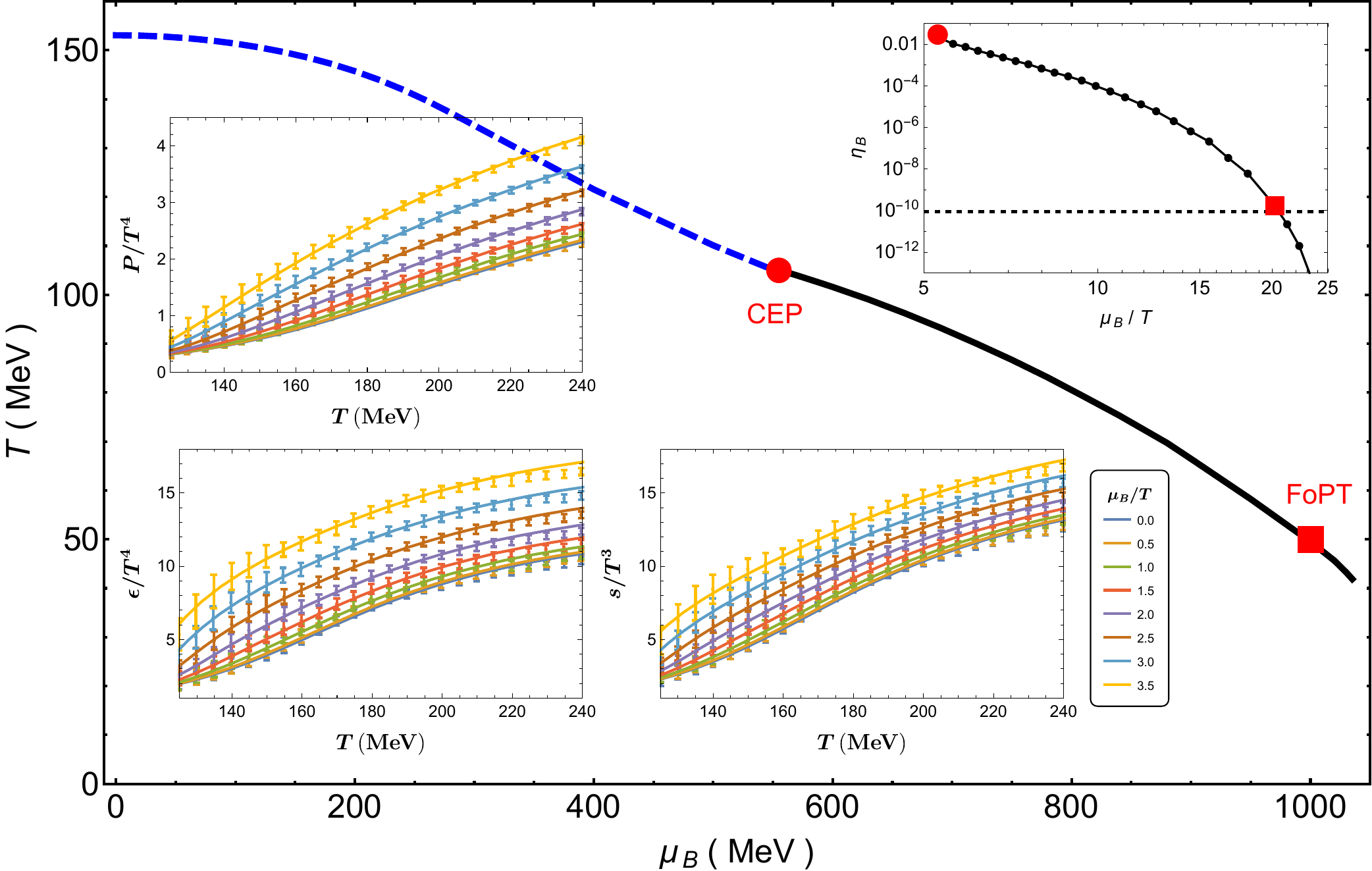}
\caption{Summary of our holographic model predictions. The QCD phase diagram is shown in the $T$-$\mu_B$ plane with the red point (CEP: critical endpoint) separating the crossover (blue dashed) and first-order (black solid) regimes. The blue dashed line is determined by the maximally increasing point of the baryon number susceptibility and the black solid one is by the free energy. The bottom-left three insets present our holographic computations of pressure $P$, energy density $\epsilon$, and entropy density $s$ compared to the latest lattice data (with error bars) within the available range $0<\mu_B/T\leq 3.5$~\cite{Borsanyi:2021sxv}, while the upper-right inset presents our model prediction on the baryon asymmetry with the red square (FoPT: first-order phase transition) singled out to match the observational value (dashed line). }\label{fig:holomodel}
\end{figure*}

While both experimental data and lattice QCD provide insights mainly within the crossover region with $\mu_B/T\leq 3.5$, we have leveraged holographic duality to establish a connection between the non-perturbative dynamics of QCD and a higher-dimensional gravity system. Our holographic model has not only demonstrated a remarkable quantitative agreement with state-of-the-art lattice QCD data for 2+1 flavors~\cite{Cai:2022omk} (see also Refs.~\cite{He:2022amv,Zhao:2022uxc,Zhao:2023gur,Cai:2024eqa} and particularly~\cite{Morgante:2022zvc}), but it also has recently exhibited consistency with experimental data from heavy ion collisions regarding baryon number fluctuations along chemical freeze-out~\cite{Li:2023mpv}. Notably, the critical endpoint (CEP) in the QCD phase diagram is located at $(T_\mathrm{CEP}=105~\text{MeV}, \mu_\mathrm{CEP}=555~\text{MeV})$, a region that is anticipated to be accessible to upcoming experimental measurements~\cite{Cai:2022omk}.
Intriguingly, our theory within the confines of the Standard Model finds that around $\mu_B=1000$ MeV, not only does the QCD phase transition become first order, but also the inferred value of $\eta_B$ aligns with cosmological observations. This presents a compelling scenario where the early universe, as described by the Standard Model, could serve as a promising source of observed GW backgrounds.

In this paper, we will use the NANOGrav 15yr data~\cite{NANOGrav:2023gor} to constrain the parameter space of the FoPT predicted by our holographic QCD model, assuming that it produces the dominant contribution to the NANOGrav signals. In particular, the Bayes parameter inferences allow us to put a strong constraint on the duration of this FoPT. Other observation constraints will be considered, including the produced primordial black holes (PBHs) and the curvature perturbations.

\textit{\textbf{Holographic model.}---}
The holographic model used to describe QCD with 2+1 flavors is represented by the following action~\cite{Cai:2022omk}
\begin{equation}\label{eq21}
  S_M =\frac{1}{2\kappa_N^2}\int \mathrm{d}^5x\sqrt{-g}\left[\mathcal{R}-\frac{1}{2}(\nabla\phi)^2-\frac{Z(\phi)}{4}F_{\mu\nu}F^{\mu\nu}-V(\phi)\right],
\end{equation}
where $A_\mu$ is the gauge field incorporating finite baryon density and $\phi$ accounts for the breaking of conformal invariance in the dual system. Alongside the effective Newton constant $\kappa_{N}^{2}$, $V(\phi)$ and $Z(\phi)$ are two independent couplings within our bottom-up model. The solution describing the hairy black hole configuration is given by
\begin{equation}\label{phiansatz}
\begin{split}
\mathrm{d}s^2=-f(r) e^{-\eta(r)} \mathrm{d}t^2+\frac{\mathrm{d}r^2}{f(r)}+r^2\mathrm{d}\boldsymbol{x_3^2}\,,\\
\phi=\phi(r),\quad A_t=A_t(r)\,.
\end{split}
\end{equation}
In this context, $\mathrm{d}\boldsymbol{x_3^2}=\mathrm{d}x^2+\mathrm{d}y^2+\mathrm{d}z^2$, and $r$ denotes the radial coordinate in the holographic setup. The AdS boundary is located as $r\rightarrow\infty$. Thermodynamic quantities such as temperature $T$, entropy density $s$, energy density $\epsilon$, and pressure $P$ can be straightforwardly derived using the standard holographic dictionary.

To encapsulate non-perturbative effects and flavor dynamics, we have employed global fitting techniques to calibrate the model parameters with state-of-the-art lattice data for (2+1)-flavors at zero net-baryon density. The explicit forms of $V(\phi)$ and $Z(\phi)$ are given by:
\begin{equation}\begin{aligned}\label{vpossi}
V(\phi)&=-12\cosh[c_1\phi]+(6 c_1^2-\frac{3}{2})\phi^{2}+{c_2\phi^{6}}\,,\\
Z(\phi) &=\frac{1}{1+c_{3}} {\text{sech}[c_4\phi^3]}+\frac{c_{3}}{1+c_{3}}e^{-c_{5} \phi}\,,
\end{aligned}\end{equation}
with $c_1=0.7100, c_2=0.0037, c_{3}=1.935, c_{4}=0.085, c_{5}=30$. Moreover, $\kappa_{N}^{2} =2 \pi({1.68})$ and the source of $\phi$ reads $\phi_s=r \phi|_{r\rightarrow\infty}=1085$ MeV that essentially breaks the conformal symmetry and plays the role of the energy scale. Further details can be found in Ref.~\cite{Cai:2022omk}.

This comprehensive framework aligns theoretical predictions with the underlying physics of the QCD phase transition and yields insights into its thermodynamic properties as shown in the phase diagram of Fig.~\ref{fig:holomodel}, where the bottom-left three insets present direction comparisons between our holographic computations on the pressure, energy density, and entropy density with respective to the latest lattice results~\cite{Borsanyi:2021sxv} only available for $\mu_B/T\leq 3.5$, while the upper-right inset presents our model prediction on the baryon asymmetry with the red square coincided with the observational value. In particular, our model predicts the phase transition between the color-neutral hadronic phase at low $T$ and small $\mu_B$ and the quark-gluon plasma at high $T$ and large $\mu_B$. The transition is a smooth crossover at small $\mu_B$ and changes into a first-order one for higher $\mu_B$. The critical point between them is at $(T_\mathrm{CEP}=105~\text{MeV}, \mu_\mathrm{CEP}=555~\text{MeV})$ which is denoted as the red point of Fig.~\ref{fig:holomodel}. 
We are particularly interested in the first-order QCD phase transition at $\mu_B=1000~\text{MeV}$ (the red square of Fig.~\ref{fig:holomodel}), which agrees with the observational constraint of a tiny baryon asymmetry, and thus provides a scenario for phase transition GWs within the Standard Model. The early state of the universe before the QCD phase transition has a relatively large baryon asymmetry $\eta_B\sim 0.1$, for which a natural mechanism for generating such a high initial baryon asymmetry is the well-established Affleck-Dine baryogenesis~\cite{Affleck:1984fy}.

For the FoPT at $\mu_B=1000$ MeV, the critical temperature $T_*=49.53$ MeV and the phase transition strength between the false ($+$) and true ($-$) vacuum reads
\begin{equation}\label{alpha}
\alpha=\frac{\theta_+-\theta_-}{3w_+}\Big{|}_{T=T_n}=\frac{\epsilon_+(T_n)-\epsilon_-(T_n)}{3w_+(T_n)}=0.33,
\end{equation}
with $\theta=\epsilon-3P$ the trace anomaly and $w=\epsilon+P$ the enthalpy. The effective number of relativistic degrees of freedom $g_\mathrm{dof}=45 s_+/(2\pi^2 T_*^3)=185$.

\begin{figure*}
\centering
\includegraphics[width=0.95\textwidth]{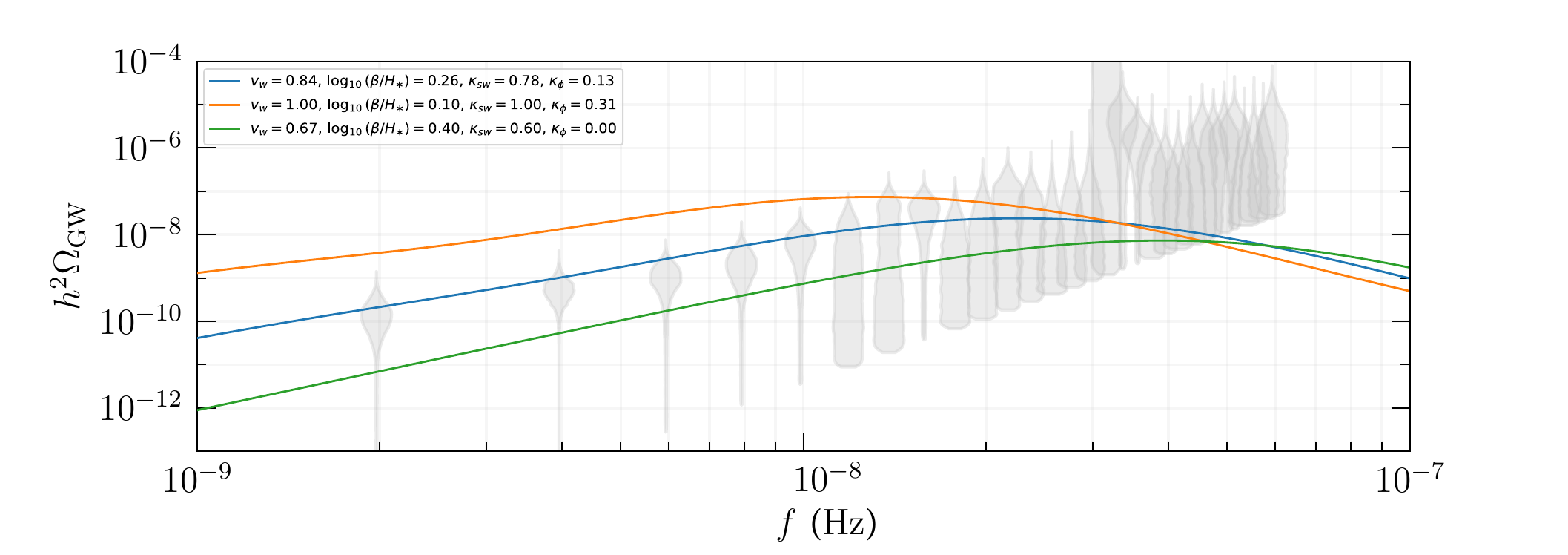}
\caption{Energy-density fraction spectra with three different sets of values for the four independent model parameters. Violin data points stand for the NANOGrav 15yr observations \cite{NANOGrav:2023gor}. }\label{fig:edsexam}
\end{figure*}

\textit{\textbf{Gravitational waves.}---}
As our holographic model predicts a FoPT around the temperature $T_*=49.53$ MeV with a strength factor $\alpha=0.33$ and the effective number of degrees of freedom $g_\mathrm{dof}=185$, the parameter space for the produced GWs shrinks down to four parameters, namely, the effective duration $\beta^{-1}$ of phase transition appeared in the combination $\beta/H_*$ with $H_*$ being the Hubble parameter at $T_*$, the terminal wall speed $v_w$ of bubble expansion, the efficiency factor $\kappa_\phi$ of converting the released vacuum energy into the wall motion, and the efficiency factor $\kappa_\mathrm{sw}$ of converting the released vacuum energy into the fluid motions. The GW spectrum from bubble wall collisions is analytically captured by the envelope approximation with the fitting formula~\cite{Weir:2017wfa,Jinno:2016vai,Huber:2008hg},
\begin{align}
h^2\Omega_\mathrm{env}=1.67\times10^{-5}&\left(\frac{100}{g_\mathrm{dof}}\right)^\frac13\left(\frac{H_*}{\beta}\right)^2\left(\frac{\kappa_\phi\alpha}{1+\alpha}\right)^2\nonumber\\
&\times\frac{0.48v_w^3}{1+5.3v_w^2+5v_w^4}S_\mathrm{env}(f),
\end{align}
where the spectral shape is of a three-section form,
\begin{align}
S_\mathrm{env}(f)=\left[c_l\left(\frac{f}{f_\mathrm{env}}\right)^{-3}+c_m\left(\frac{f}{f_\mathrm{env}}\right)^{-1}+c_h\left(\frac{f}{f_\mathrm{env}}\right)\right]^{-1}
\end{align}
with $c_l=0.064$, $c_h=0.48$, and $c_m\equiv 1-c_l-c_h$, and the peak frequency is given by
\begin{align}
f_\mathrm{env}=1.65\times10^{-5}\,\mathrm{Hz}&\left(\frac{g_\mathrm{dof}}{100}\right)^\frac16\left(\frac{T_*}{100\,\mathrm{GeV}}\right)\nonumber\\
&\times\frac{0.35(\beta/H_*)}{1+0.069v_w+0.69v_w^4}.
\end{align}
The GW spectrum from fluid motions is dominated by sound waves~\cite{Hindmarsh:2013xza,Hindmarsh:2015qta,Hindmarsh:2017gnf,Hindmarsh:2016lnk,Hindmarsh:2019phv,Guo:2020grp,Cai:2023guc} fitted by numerical simulations~~\cite{Hindmarsh:2013xza,Hindmarsh:2015qta,Hindmarsh:2017gnf} as
\begin{align}
h^2\Omega_\mathrm{sw}=2.65\times10^{-6}&\left(\frac{100}{g_\mathrm{dof}}\right)^\frac13\left(\frac{H_*}{\beta}\right)\left(\frac{\kappa_\mathrm{sw}\alpha}{1+\alpha}\right)^2\nonumber\\
&\times\frac{7^{7/2}v_w(f/f_\mathrm{sw})^3}{(4+3(f/f_\mathrm{sw})^2)^{7/2}}\Upsilon
\end{align}
with the peak frequency given by
\begin{align}
f_\mathrm{sw}=1.9\times10^{-5}\,\mathrm{Hz}\left(\frac{g_\mathrm{dof}}{100}\right)^\frac16\left(\frac{T_*}{100\,\mathrm{GeV}}\right)\left(\frac{1}{v_w}\right)\left(\frac{\beta}{H_*}\right)\ ,
\end{align}
where the spectral shape at low frequencies can be analytically modeled~\cite{Cai:2023guc} as forced collisions of sound shells during bubble percolations, while the spectral shape at high frequencies can be analytically modeled~\cite{Hindmarsh:2016lnk, Hindmarsh:2019phv} as free collisions of sound shells long after bubble percolations. Here, the suppression factor $\Upsilon\equiv1-(1+2\tau_\mathrm{sw}H_*)^{-1/2}$~\cite{Guo:2020grp} accounts for the finite lifetime of sound waves from the onset timescale of turbulences, $\tau_\mathrm{sw}H_*\approx(8\pi)^{1/3}v_w/(\beta/H_*)/\bar{U}_f$ with the root-mean-squared fluid speed $\bar{U}_f^2=3\kappa_\mathrm{sw}\alpha/[4(1+\alpha)]$. The contribution from magnetohydrodynamic turbulences is neglected since at most $5\%-10\%$ of fluid motions is turbulent~\cite{Hindmarsh:2015qta,Caprini:2015zlo}.

\begin{figure}
\includegraphics[width=\columnwidth]{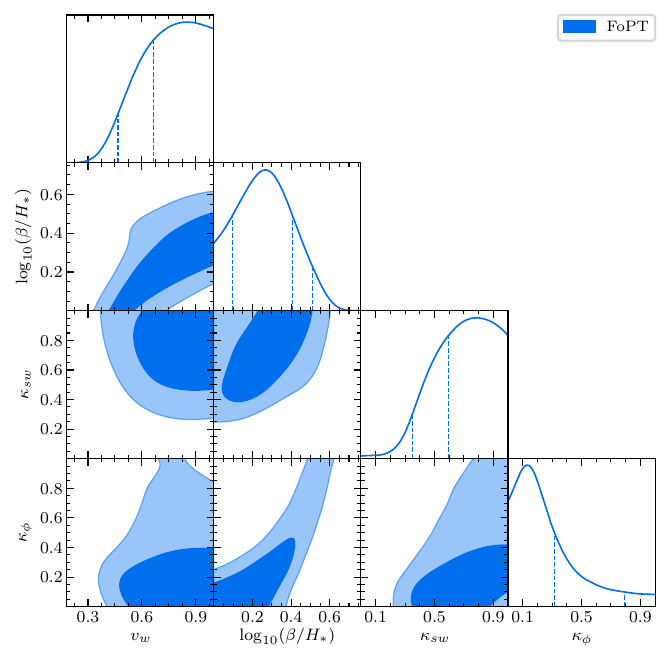}
\caption{Posteriors of the four independent model parameters inferred from the NANOGrav 15-year data release \cite{NANOGrav:2023gor}.}\label{fig:posteriors}
\end{figure}

\textit{\textbf{PTA constraints.}---} Our holographic model has already fixed three parameters ($T_*, \alpha, g_\mathrm{dof}$) but is left with four parameters ($\beta/H_*, v_w, \kappa_\phi, \kappa_\mathrm{sw}$), which are all independent parameters \textit{in practice} as argued shortly below. For a bag equation of state (EoS), the efficiency factor $\kappa_\mathrm{sw}$ of fluid motions can be determined as a function~\cite{Espinosa:2010hh} of the strength factor $\alpha$ and wall speed $v_w$, but it eventually becomes model-dependent when going beyond the bag EoS~\cite{Giese:2020znk, Giese:2020rtr,Wang:2020nzm,Wang:2022lyd,Wang:2023jto}. For our holographic model, the sound speeds in the false and true vacua can be calculated as $c_+^2=0.15$ and $c_-^2=0.14$, respectively, which deviate significantly from the bag EoS with the sound speed $c_s^2=1/3$. Hence, we will treat $\kappa_\mathrm{sw}$ as an independent parameter. The determination for the wall speed $v_w$ is even more model-dependent~\cite{Moore:1995ua,Moore:1995si,Konstandin:2014zta,Laurent:2020gpg,Laurent:2022jrs} (see, however, the recent attempts of model-independent approaches~\cite{Ai:2021kak,Li:2023xto} from local equilibrium and strong coupling~\cite{Bea:2021zsu,Janik:2021jbq}, respectively), and hence the wall speed $v_w$ is also treated as an independent parameter. As for the efficiency factor $\kappa_\phi$ of wall collisions, it always admits an extra dependence on the leading-order friction term~\cite{Ellis:2019oqb,Ellis:2020nnr,Cai:2020djd} whenever the GWs are dominated by wall collisions or fluid motions~\cite{Cai:2020djd,Lewicki:2022pdb}. Therefore, $\kappa_\phi$ is also model-dependent and treated as an independent parameter as well. Last, $\beta/H_*$ is an independent parameter of FoPT on its own (see, however, Ref.~\cite{Ares:2021nap} from a holographic computation).

Following the approach of Ref.~\cite{NANOGrav:2023hvm}, we adopt the publicly available \texttt{PTArcade} \cite{andreamitridate_2023,Mitridate:2023oar} \footnote{\url{https://andrea-mitridate.github.io/PTArcade/getting_started/}}, which provides a wrapper of \texttt{enterprise} \cite{enterprise,enterprise-ext} \footnote{\url{https://github.com/nanograv/enterprise}}, to perform Bayes parameter inferences and obtain the parameter region allowed by the NANOGrav 15-year data \cite{NANOGrav:2023gor}. 
We set flat priors for all the four independent model parameters in their allowing ranges within $\log_{10}(\beta/H_{\ast})\in[0,3]$, $v_{w}\in[0,1]$, $\kappa_{\phi}\in[0,1]$, and $\kappa_{\mathrm{sw}}\in[0,1]$, respectively. Here, the upper bound for $\beta/H_*$ is conservatively chosen with a large number.
In Fig.~\ref{fig:edsexam}, we depict the GW energy-density fraction spectra, given three different sets of values for these independent parameters. 
Performing Bayes analysis, we obtain the posteriors of these parameters that are shown in Fig.~\ref{fig:posteriors}. 
Correspondingly, the median values and uncertainties of these parameters are inferred to be $\log_{10}(\beta/H_\ast)=0.26_{-0.16}^{+0.14}$, $v_{w}=0.84_{-0.17}^{+0.16}$, $\kappa_{\phi}=0.13_{-0.13}^{+0.18}$, and $\kappa_{\mathrm{sw}}=0.78_{-0.18}^{+0.22}$ at $68\%$ confidence level. Furthermore, we plot the blue solid curve in Fig.~\ref{fig:edsexam} from the peak value of one-dimensional posterior for each parameter after marginalized over all the other parameters as shown in Fig.~\ref{fig:posteriors}. The curve seems to be capable of fitting nicely with the NANOGrav 15yr data.

\begin{figure}
\includegraphics[width=\columnwidth]{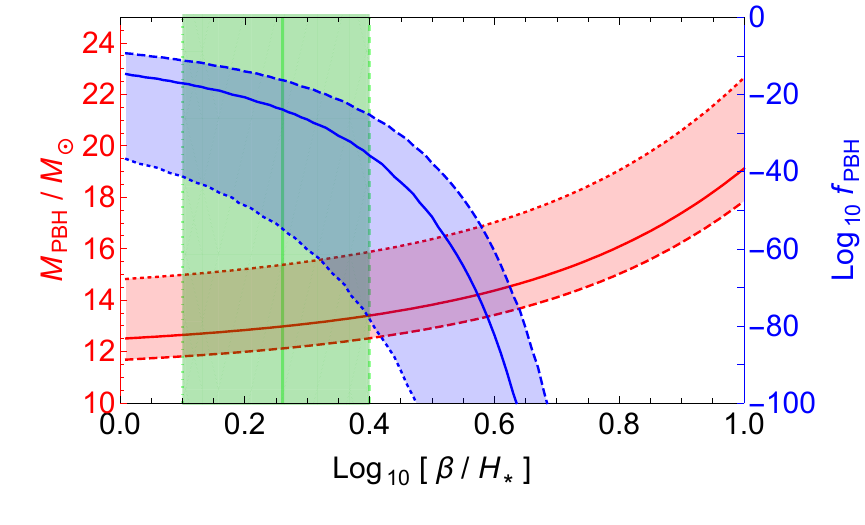}
\caption{The constraints on the PBH mass (red) and abundance (blue) from the delayed decay mechanism given the PTA constraints on $\log_{10}(\beta/H_*)$ (green) and $v_w$, where the dotted and dashed curves are obtained from the corresponding $1\sigma$ lower and upper bounds of $v_w$, respectively.}\label{fig:MPBHfPBH}
\end{figure}

\textit{\textbf{Other constraints.}---} As a general property of any FoPT, the vacuum decay process is not simultaneous all over the space, and hence there is always a non-vanishing chance to find Hubble-scale regions where the vacuum decay progress falls behind their ambient regions~\cite{Guth:1982pn}. Since the false vacuum energy can never be diluted away while the radiations would with the Hubble expansion of our Universe, the total energy density in these delayed-decay regions would gradually accumulate their density contrasts to enhance the curvature perturbations~\cite{Liu:2022lvz} or even form PBHs~\cite{Liu:2021svg} when exceeding the PBH threshold. See also Refs.~\cite{Gouttenoire:2023naa,Baldes:2023rqv} for recent improved treatments on Refs.~\cite{Lewicki:2023ioy,Kodama:1982sf}. Therefore, there are always accompanying constraints other than GWs from PBHs and curvature perturbations. Following the same procedure of Ref.~\cite{Liu:2021svg} (see also Ref.~\cite{He:2022amv}), we can calculate the associated PBH formations as shown in Fig.~\ref{fig:MPBHfPBH}, where the PBH mass (red) and abundance (blue) are obtained for given $\beta/H_*$ and $v_w$ with their $1\sigma$ uncertainties expanded between the dotted (lower bound) and dashed (upper bound) lines. All the other parameters are fixed by our holographic model. In particular, the sound speed $c_+^2=0.15$ in the false vacuum analytically fixes the PBH threshold $\delta_c=[3(1+w)/(5+3w)]\sin^2[\pi\sqrt{w}/(1+3w)]=0.35$~\cite{Harada:2013epa}. It is easy to see that the produced PBH mass is roughly between $\sim12-16\,M_\odot$ and the PBH abundance $f_\mathrm{PBH}\lesssim10^{-10}$ is so tiny that the current observational constraint~\cite{Chen:2021nxo} in this mass range can still be evaded. Therefore, the overlapping region from the joined red, blue, and green shaded regions is still allowed by the current PBH observations.

However, such a low-scale phase transition at $T_*=49.53$ MeV with this large $\alpha=0.33$ and small $\beta/H_*\sim1.3-2.5$ might be disfavoured by the current constraints on curvature perturbations~\cite{Liu:2022lvz} from the ultracompact minihalo (UCMH) abundance~\cite{Clark:2015sha,Clark:2015tha} but depending on the choice of the window function~\cite{Cai:2024nln}. This is evident from Fig.~2 of Ref.~\cite{Liu:2022lvz} that the parameter space of a phase transition at $T_*=49.53$ MeV with a fixed $\alpha=0.33$ has been ruled out for all $\beta/H_*\lesssim100$ by UCMH observations~\cite{Clark:2015sha,Clark:2015tha}. Nevertheless, this conclusion might be sensitive to the choice of the window function~\cite{Elor:2023xbz} as shown in Ref.~\cite{Cai:2024nln} for a top-hat window function with an illustrative choice of the cutoff, and the corresponding constraints on the FOPT from various curvature perturbation observations differ significantly from that in Ref.~\cite{Liu:2022lvz}. Note that a small $\beta/H_*$ is also argued to be disfavored from the holographic side but only at the probe limit~\cite{Chen:2022cgj}. Therefore, our holographic model at the benchmark point with $\mu_B=1000$ MeV at least fits into the current constraints of the combined regions from both PTA and PBH observations, while the consistency with the curvature-perturbation constraints would call for future study with more dedicated treatments. Furthermore, the large chemical potential would require a little inflation~\cite{Boeckel:2009ej,Boeckel:2011yj} during the QCD phase transition to achieve the observed baryon asymmetry, which will also be reserved for future study.

\textit{\textbf{Conclusions and discussions.}---}
The cosmological QCD phase transition is still a myth to both communities from particle physics and nuclear physics, and whether it is of the first order can be tested by stochastic GW backgrounds possibly detectable from PTA observations. Recent observations of a low-frequency GW background from NANOGrav, EPTA, PPTA, and CPTA provide a promising opportunity to test various first-order QCD phase transition models, in particular, our holographic model aligned quantitatively with lattice QCD data, which is strongly constrained by the NANOGrav 15yr data, especially for its phase-transition duration parameter. 

All these constraints, along with the parameters fixed already by the holographic model, can be transformed into constraints on the associated PBH formations by the delayed decay mechanism, and the produced PBH abundance is still observationally allowed, even though the constraints on the induced curvature perturbations might be sensitive to the choices of window functions. Although the QCD phase transition from our holographic scenario is consistent with the recent GW background~\cite{NANOGrav:2023gor} as shown with the blue curve in Fig.~\ref{fig:edsexam}, other possibilities would also exist. In particular, there can be other GW sources that also contribute to the GW background.

\begin{acknowledgments}
We thank Rong-Gen Cai, Qing-Xuan Fu, Yuan-Xu Wang, Hong-An Zeng, and Zhi-Chao Zhao for the valuable discussions. This work is supported by the National Key Research and Development Program of China Grant No. 2020YFC2201501, No. 2021YFC2203004, and No. 2021YFA0718304,
the National Natural Science Foundation of China Grants No. 12075101, No. 12235016, No. 12122513, No. 12075298, No. 12175243, No. 12105344, No. 11991052, No. 12235019, and No. 12047503, the Fundamental Research Funds for the Central Universities, the Max Planck Partner group research grant, and the Science Research Grants from the China Manned Space Project with No. CMS-CSST-2021-B01.
\end{acknowledgments}


\bibliography{ref}

\end{document}